# A Hybrid Image Cryptosystem Based On OMFLIP Permutation Cipher

G.Sudheer[1] and B.V.S.Renuka Devi[2]

[1]Dept. of Mathematics, GVP College of Engineering for Women, India
```
g.sudheer@gvpcew.ac.in
g_sudheer@hotmail.com
```
[2] Dept. of ECE, GVP College of Engineering for Women, India
```
bvsrenuka@gvpcew.ac.in
bhamidipati.renuka@gmail.com
```

*Abstract*

*The protection of confidential image data from unauthorized access is an important area of research in network communication. This paper presents a high-level security encryption scheme for gray scale images. The gray level image is first decomposed into binary images using bit scale decomposition. Each binary image is then compressed by selecting a good scanning path that minimizes the total number of bits needed to encode the bit sequence along the scanning path using two dimensional run encoding. The compressed bit string is then scrambled iteratively using a pseudo-random number generator and finally encrypted using a bit level permutation OMFLIP. The performance is tested, illustrated and discussed.*

## 1. Introduction

Advances in computer and communication technologies have facilitated the emergence of efficient systems for delivery of a wide range of multimedia data. In particular, the rapid development of internet communication has led to the wide spread deployment of digital image services. Security is an important issue in the transmission and storage of digital images and encryption is a means of ensuring security. It is well known that images are different from text data in many aspects such as large size of data, high unwanted redundancy and correlation. Therefore it is not always a good idea to use traditional cryptosystems (such as DES, AES, RSA etc.) on the image directly. During the last decade many encryption schemes for image data have been proposed. In many of these schemes, secret permutations have been used to shuffle the positions of pixels (and/or pixel bits), which is an effective and easy way to make the cipher image look chaotic [1]. The security of permutation-only multimedia ciphers has been extensively studied and their vulnerability against cipher text-only attacks, known/chosen plaintext attack has been a source of concern in the effective deployment of the proposed cryptosystems. The insecurity against these shortcomings can be solved in practice by combining permutation operations with substitution operations, using complicated permutations, dynamically updating the permutations, etc [2]. The design of secure image cryptosystems can be bolstered by combining permutation operations with other encryption mechanisms. In this context, new permutation instructions, PPERM, GRP, CROSS, OMFLIP [3] will be useful for faster cryptography and quick multimedia processing. In this paper OMFLIP, a permutation instruction using enhanced Omega-Flip interconnection network theory is combined with SCAN methodology based scan patterns to develop an image encryption technique. The transposition-substitution technique is first applied on the compressed bit planes of gray scale image and is followed with OMFLIP permutation operation to provide confusion and diffusion to the encrypted data.

The rest of the paper is organized as follows: Section 2 surveys some related image cryptosystems and their salient features. The proposed encryption scheme together with its different stages is presented in section 3. The relevant security parameters and the decryption scheme are outlined in section 4. The experimental results of the proposed image cryptosystem are presented and discussed in section 5 and is followed with concluding remarks in section 6.

## 2. Related Image Cryptosystems

In order to transmit secret images to other people a variety of image encryption methods have been proposed in recent years. They include permutation based methods, chaos-based methods, vector quantization based methods and other miscellaneous methods. [4] and [5] developed a picture data encryption scheme for binary images using scan patterns. The scheme converts a 2D image into a 1-dimensional list and employs a scan language [6] to describe the converted result. [7] presented a new method for binary image encryption using modified scan language. By putting different scan patterns at the same level in the scan tree structure and employing a two dimensional run encoding (2DRE) technique, they claimed that their method can encrypt large binary images with higher security. [1] found vulnerabilities in the scheme proposed by [7]. They have shown that the scheme could be broken with some chosen pairs of plain and cipher image. [8] presented a new methodology which performs both lossless compression and encryption of binary and gray scale images. At the core of the compression method is the algorithm which determines a near optimal scanning path that minimizes the total number of bits needed to represent the encoded scanning path and the encoded bit sequence along the scanning path. This framework was used again to present an iterated product cipher that is based on repeated and intertwined application of permutation and substitution operations [9]. Meanwhile [10] presented an improved encryption scheme for binary image. The improvement was on [7] scheme utilizing some of the suggestions given by [1]. It is observed that all scan-based image and video encryption schemes can be divided into two generations: the first generation uses secret permutations to encrypt plain images and the second combines secret permutations and the substitution algorithm. The first generation is not secure enough against known-plain text and chosen-plain text attacks while the second can enhance the security of the first generation [11]. In all scan-based encryption schemes, the secret permutations are generated from 13 different scan patterns and 24 different partition patterns, where the scan words serve as the secret key. The scan patterns are combined to generate a scan order of the image [9]. The present work aims at utilizing the SCAN methodology based patterns for optimal scan order selection.

## 3. The Proposed Scheme

An image can be viewed as an arrangement of bits, pixels and blocks. The perceivable information present in an image due to correlations among the bits, pixels and blocks can be reduced using permutation techniques. As bit level permutation operations are very important from both an architectural and cryptographic points of view, the present paper proposes an encryption scheme for gray level images combining scan language based patterns, two dimensional run-encoding (2DRE) and OMFLIP permutation operation. As permutations are new requirement for fast processing of digital multimedia information with sub word-parallel instructions, more commonly known as multimedia instructions, the choice of OMFLIP operation for the proposed image cryptosystem seems appropriate. The hardware complexity of the permutation functional unit corresponding to the OMFLIP instruction is very low and the memory requirement is also small compared to other permutation operations like GRP, PPERM, CROSS, etc. [3], making it ideal for encryption.

The proposed scheme consists of the following stages
  i. Use Bit-level decomposition to obtain bit planes of a gray level image.
  ii. Use two dimensional run encoding (2DRE) to compress the binary image (each bit plane) after selecting an optimal scanning path utilizing SCAN methodology based patterns.
  iii. Scramble the compressed bit string using a pseudorandom number generator.
  iv. Utilize randomly generated control bits and OMFLIP operation to encrypt the scrambled bit strings.

The critical stage of employing control bits and OMFLIP operations makes the image cryptosystem less vulnerable to attacks.

## 3.1. Bit-plane decomposition

A gray scale image consisting of a maximum of N gray levels can be decomposed into $\log_2 N$ bit-planes, where N is a power of 2. Let $X(m, n)$ be a $2^L$ –gray image of size M X N and $X_P(m,n)$ be a pixel value of the image($1 \leq m \leq M$, $1 \leq n \leq N$). Then the bit decomposition of the image is described as

$$X^l(m, n) = B^l(X_P(m, n)) \qquad (1)$$

where $B^l(.)$ represents the operator of the bit decomposition, and $X^l(m,n) \in \{0,1\}$. A gray scale image is the sum of the bit planes weighted by their respective grayness. The composition is given by

$$X(m,n) = \sum_{l=0}^{L-1} X^l(m,n) 2^l \qquad (2)$$

In essence a bit plane is a set of bits having the same position in the respective binary number representation of the gray scale values. Thus a 256 gray level image is decomposed into eight bit planes (binary images) $b_0$, $b_1$, $b_2$, $b_3$, $b_4$, $b_5$, $b_6$ and $b_7$ starting from the least significant bit plane to the most significant bit plane. On each bit plane, the two dimensional run encoding, scan pattern encryption and OMFLIP permutation are carried out.

## 3.2. Two dimensional run encoding (2DRE) based on optimal scan order

The SCAN [6] is a formal language based two dimensional spatial accessing methodology that can generate a wide variety of scanning paths easily. As the compression of a binary image is done by specifying a scanning path and the bit sequence along that scanning path, an optimal scanning path that yields small number of large segments of 0s and 1s is chosen. Along the chosen scan order, 2DRE is carried out. It is a simple technique used to compress data. The basic idea in 2DRE is to count the number of times the same bit repeats successively in the scan order. The first bit of the scan order in an image is recorded and the successively repeated counts according to the scan order are saved. The process is recursively carried out for the other bits. Each bit plane is thus encoded as a one dimensional string. This is the compressed image.

### 3.3. Scrambling

The 2DRE of a binary image along a good scanning path results in a compressed string of runs. The maximum run length gives an idea of the minimum number of bits required to represent each element of the runs string. The runs string is encoded to a binary form and grouped into blocks for encryption using a rearrangement of bits. The number of bits ($x \geq 3$) to be grouped together into a block is decided by testing if the total length of the compressed bit string is divisible by x. A sequence of x! patterns are generated and a chosen order of patterns is used iteratively to change the position of bits in all the blocks. On the scrambled bit string OMFLIP permutation is applied to render statistical analysis of the encrypted bit string useless.

### 3.4. OMFLIP

The OMFLIP operation is basically a concatenation of two permutation stages – an omega stage and a flip stage. In an omega or flip stage, w input bits are divided into w/2 pairs. The two bits in a pair are mapped to two output positions, the destination order being determined by a single control bit. At the input of an omega stage, bits i and (i+w/2), $0 \leq i < w/2$ form a pair and they are mapped to the two bit positions 2i and (2i+1). At the input of a flip stage, bits 2i and (2i+1), $0 \leq i < w/2$, form a pair which is mapped to positions i and i+w/2. A flip stage can be viewed as the inverse of omega stage. The OMFLIP operation $Z = X \lozenge_{(a0,a1)} Y$ uses two stages in an omega-flip network to permute the data bits X and Y specifying the control bits for the two stages. The subscript (a0, a1) represents a two bit encoding that specifies which stages are used, they could be (omega, omega), (flip, flip), (omega, flip) or(flip, omega). The programmatic definition of OMFLIP is given in the appendix.

Using the OMFLIP operation together with the control bits, we encrypt the compressed bit string. This is carried out for the eight compressed bit planes changing the control bits for each bit plane. After encryption using the permutation operation, we end up with eight binary sequences of encrypted data. This is used for final transmission to the intended receiver with the required security keys.

The following are the security keys in the proposed encryption method.
   i.   SCAN order.
   ii.  The maximum run length of each compressed bit plane.
   iii. The control bit vector used for the OMFLIP operation.
   iv.  The sequence of transmission of bit planes.

The transmission can be carried out as binary sequences or as images. The binary sequence files can be further encrypted using asymmetric key cryptography for safe transmission.

## 4. Decryption

At the receiver's end, the sequence of bit planes is deciphered. Using the received control bits for the corresponding bit plane, the reverse OMFLIP operation is carried out after which the reverse scan process is carried out by the scrambling patterns. Further using the maximum run length in 2DRE, the reverse 2DRE is carried out. Once the correct runs are obtained utilizing the scan order, the corresponding bit planes are retrieved.

The decryption phase consists of the following stages
   i.   Reverse OMEGA FLIP
   ii.  Reverse scan process based on scrambling patterns
   iii. Reverse 2DRE to decompress

iv. Retrieve the bit planes using the SCAN order and combine them to obtain the gray scale image.

All the four stages in the decryption process are the reverse of the encryption stages and hence the detailed methodology is not outlined.

## 5. Results and discussions

The proposed image cryptosystem was implemented in software using MATLAB 7.0. The gray scale images considered were of size 128 × 128 with 8 bits/pixel. Lena, Cameraman and Mandrill tiff images were used in the tests. The tests were run on a Desktop PC with Intel Core 2 Duo Processor having 2 GB RAM. The security of our image cryptosystem can be analysed as follows. Consider a $2^n \times 2^n$ size image ($2 \leq n \leq 9$) with 8 bits per pixel. The bit plane decomposition method splits the image into 8 bit planes, each bit plane being of size $2^n \times 2^n$. An optimal scan order based on SCAN methodology is chosen for compression of each bit plane. The scan orders act as encryption keys. For an n × n array there are (n×n)! scanning orders possible. The SCAN methodology gives us a wide variety of scan orders and a good scanning order is chosen. Then along the scan order the 2DRE is carried out. Assuming a maximum compression ratio of r, the total bits of each compressed bit plane is $\frac{2^n \times 2^n}{r}$. The compressed bit plane is then encrypted using scrambling patterns. If the maximum run length of a bit plane is l and x bits are needed to represent that element (note that x is so chosen that the total length of compressed bit string is divisible by x), then the elements are grouped by x bits and there are $\frac{2^n \times 2^n}{r.x}$ groups to be encrypted. As x! scan patterns are generated the number of possible combinations for an attacker is $(x!)^{\frac{2^n \times 2^n}{r.x}}$. To ensure confusion and diffusion of the cipher image, the OMFLIP permutation operation is performed on the bits after scrambling the bits. An n-input omega network has log n identical omega stages. An n-input flip network being the exact mirror image of an n-input omega network also has $\log n$ identical flip stages. Based on control bit, the encrypted bit sequence is permuted. The scrambled bit sequence was checked for correlation against the OMFLIP permuted bit sequence. The two bit sequences were found to be least correlated. The correlation values of the two encrypted bit sequences in the most significant bit planes for different images is shown in table 1. There is negligible correlation between the two stages of encryption. In addition we applied another permutation cipher called GRP [12] on scrambled bit sequence. The results of correlation between the bit strings obtained after stage 3 and GRP operation are shown in table 2. Compared to GRP bit level permutation we found OMFLIP permutation to be better in achieving the desired security.  A comparison of the properties of GRP and OMFLIP operations [3] made us choose OMFLIP over GRP. The security of encryption is further enhanced as each bit plane uses a different scan pattern, control bits and may use different length blocks for scrambling.  Therefore even for a brute force attack the key-space is exhaustively large making the attack infeasible. The finally OMFLIP encrypted bit sequence was tested for statistical randomness using the binary entropy function

$$H(p) = -p * \log p - (1-p) * \log(1-p) \qquad (3)$$

where p is the percentage of 1's (or of 0's) was used as the measure. The results are tabulated in table 3.

Moreover there is high key sensitivity as it is difficult to obtain the original image from cipher bit strings even if there is a slight difference in the control bits used in the OMFLIP operation. Moreover the order of transmission of the bit plane streams adds to the security. For an image of size $64 \times 64$ or more the control bits used in the OMFLIP operation is of minimum length 300. We found that a change in 3 or more positions of the control bits was making decryption incorrect. This means that the key is highly sensitive.

A linear and differential cryptanalysis of permutation instructions such as DDR, OMFLIP and GRP was carried out in [12], [13] and they opined that the differential properties of DDR and OMFLIP seem to complement each other and one can expect good security enhancement by combining them. In the present work scan patterns and OMFLIP were combined to achieve greater security.

## 6. Conclusions

A hybrid image cryptosystem combining the features of bit level permutation cipher OMFLIP for faster cryptography and SCAN methodology based patterns for efficient compression is proposed in the present paper. The proposed scheme consists in applying the compression-encryption methodology to each bit plane of a gray scale image. SCAN methodology based scan order and 2DRE compression compresses the two dimensional bit plane into a one dimensional bit string. The bit strings are then grouped into blocks and scrambled. The scrambled bit strings are permuted based on control bit using the OMFLIP operation. The control bits are randomly generated using a seed and they differ from bit plane to bit plane. The main features of the proposed scheme are key dependent permutation, variable length keys, large key space and combines encryption with compression.

## 7.Tables:

Table 1. Correlation between bit strings obtained after stage 3 and stage 4 of encryption with OMFLIP

| Image | Bit plane8 | Bit plane7 | Bit plane6 |
|---|---|---|---|
| **Lena** | 0.0035585 | -0.010236 | 0.0066928 |
| **Cameraman** | -0.014545 | -0.0060899 | 0.0044065 |
| **Mandrill** | 0.0069122 | -0.0045668 | -0.012675 |

Table 2. Correlation between bit strings obtained after stage 3 and stage 4 of encryption with GRP

| Image | Bit plane8 | Bit plane7 | Bit plane6 |
|---|---|---|---|
| **Lena** | -0.0142776 | 0.0133210 | -0.0014794 |
| **Cameraman** | 0.0048849 | -0.0199478 | 0.0023038 |
| **Mandrill** | 0.0216418 | -0.0072855 | 0.0233466 |

Table 3.Entropy values of the cipher bit streams

| **Image** | Bit plane $b_7$ | Bit plane $b_6$ | Bit plane $b_5$ | Bit plane $b_4$ | Bit plane $b_3$ | Bit plane $b_2$ | Bit plane $b_1$ | Bit plane $b_0$ |
|---|---|---|---|---|---|---|---|---|
| **Lena** | 0.892 | 0.8611 | 0.8386 | 0.8091 | 0.7955 | 0.8792 | 0.8765 | 0.8770 |
| **Camerman** | 0.615 | 0.5972 | 0.7944 | 0.7791 | 0.8325 | 0.8924 | 0.7924 | 0.8785 |
| **Mandrill** | 0.7770 | 0.8249 | 0.8102 | 0.8826 | 0.8764 | 0.8845 | 0.8780 | 0.8788 |

# 8. Appendix

**OMFLIP** $\Diamond_{(a0,a1)}$

Step 1: length of input compressed bit string=w
Input compressed bit string=in
Output encrypted bit string=out
Control bit to select omega or flip stage ($a_0$ or $a_1$)=c
lim=$\lceil w/2 \rceil$

Step 2: if c=0 (omega stage) go to step 3
Else if c=1(flip stage) go to step 7

Step 3: out(2i)=in(i)
out(2i+1)=in(i+lim)

Step 4: if i=lim and w is even
out (1)=in(i+lim)

Step 5: if i=lim and w is odd
out(1)=in(lim)

Step 6: repeat steps 3,4 and 5 for i=1 to lim

Step 7: end

Step 8: out(i)=in(2i)
out(i+lim)=in(2i+1)

Step 9: if i=lim and w is even
out(i+lim)=in(1)

Step 10: if i=lim and w is odd
out(lim)=in(1)

Step 11: repeat steps 8, 9 and 10 for i=1 to lim